\documentclass[rnote]{aa} 
\usepackage{graphicx}
\usepackage{txfonts}
\usepackage{natbib}
\bibpunct{(}{)}{;}{a}{}{,}
\newcommand{\ltsimeq}{\raisebox{-0.6ex}{$\,\stackrel
        {\raisebox{-.2ex}{$\textstyle <$}}{\sim}\,$}}

\newcommand{\chemone}{\raisebox{0.03cm}{$-$}} 

\begin{document}
   \title{In situ apparatus for the study of clathrate hydrates relevant to solar system bodies using synchrotron X-ray diffraction and Raman spectroscopy}

   \titlerunning{In situ apparatus for clathrate hydrates}

   \author{Sarah J. Day\inst{1,2}\thanks{email: sarah.day@diamond.ac.uk}
          \and S. P. Thompson\inst{1}
	  \and A. Evans\inst{2}
	  \and J. E. Parker\inst{1} }
   
   \authorrunning{S. J. Day et al.}
   
   \institute{Diamond Light Source, Harwell Science and Innovation Campus,
	 Didcot, Oxfordshire, OX11 0DE, UK \\      
         \and Astrophysics Group, Keele University, Keele, Staffordshire, ST5 5BG, UK}

   \date{Version of 06-01-2015}

\abstract{Clathrate hydrates are believed to play a significant role in
          various solar system environments,
	  e.g. comets, and the surfaces and interiors of icy
	  satellites. However, the structural factors governing their formation and dissociation are poorly understood.} 
         {We demonstrate the application of a high pressure gas cell, combined with variable temperature non-contact cooling and fast, time-resolved data collection, to the in situ study of clathrate hydrates under conditions relevant to solar system environments.} 
	 {Clathrates formed and processed within the sample cell are monitored in situ using time-resolved synchrotron X-ray powder diffraction and laser Raman spectroscopy.} 
	 {X-ray diffraction allows the formation of clathrate hydrates to be observed as CO$_{2}$ gas is applied to ice formed within the cell. Complete conversion is obtained by annealing at temperatures just below the ice melting point. A subsequent rise in the quantity of clathrate is observed as the cell is thermally cycled. Four regions between 100 -- 5000 cm$^{-1}$ are present in the in situ Raman spectra that carry features characteristic of both ice and clathrate formation.}
         {This novel experimental arrangement is well suited to studying clathrate hydrates over a wide range of temperature (80~K -- 500~K) and pressure (1--100~bar) conditions relevant to solar system bodies and can be used with a variety of different gases and starting aqueous compositions (e.g. saline solutions). We propose the increase in clathrate formation observed during thermal cycling may be due to the formation of a quasi liquid-like phase that forms at temperatures below the ice melting point, but which allows either easier formation of new clathrate cages, or the retention and delocalisation of previously formed clathrate structures, possibly as amorphous clathrate. The structural similarities between hexagonal ice, the quasi liquid-like phase, and crystalline CO$_{2}$ hydrate mean that differences in the Raman spectrum are subtle; however, all features out to 5000 cm$^{-1}$, when viewed together, are diagnostic of clathrate structure.}


   \maketitle
%

\section{Introduction \label{introclath}}

Clathrate hydrates are compound structures consisting of water-ice in which
water molecules form cage structures enclosing gas molecules (referred to as the
host and guest molecules, respectively). They are stable only in specific
temperature and pressure regimes that depend on the size of the guest species,
and the structure of the ice cages themselves \citep[see][for
reviews]{sloan98,sloan03}. Clathrates can form three independent structures that are determined by the size
and shape of the cages formed. Structure~I (sI) clathrates form a cubic crystal structure of space group {\it{Pm-3n}} and are characterised
by two cage types,  namely $5^{12}$ (i.e. constructed of 12 pentagonal faces) and the larger $5^{12}6^2$ (i.e. 12~pentagonal
and  2~hexagonal faces). They consist of two small cages to every six large cages, and
typically host comparatively large guest molecules such as CO$_2$ and CH$_4$. Structure~II (sII) clathrates incorporate smaller molecules such as O$_2$ and N$_2$, forming
sixteen small $5^{12}$ cages and eight large $5^{12}6^4$
cages. Structure~H (sH) clathrates are less common,
consisting of three small cages, two medium cages, and one large cage per unit
cell, and require two guest species to fill the cages in order to remain stable. 
%

\begin{figure*}
\setlength{\unitlength}{1cm}
\begin{picture}(12.0,6.5)
\put(0.8,-1.1){\includegraphics{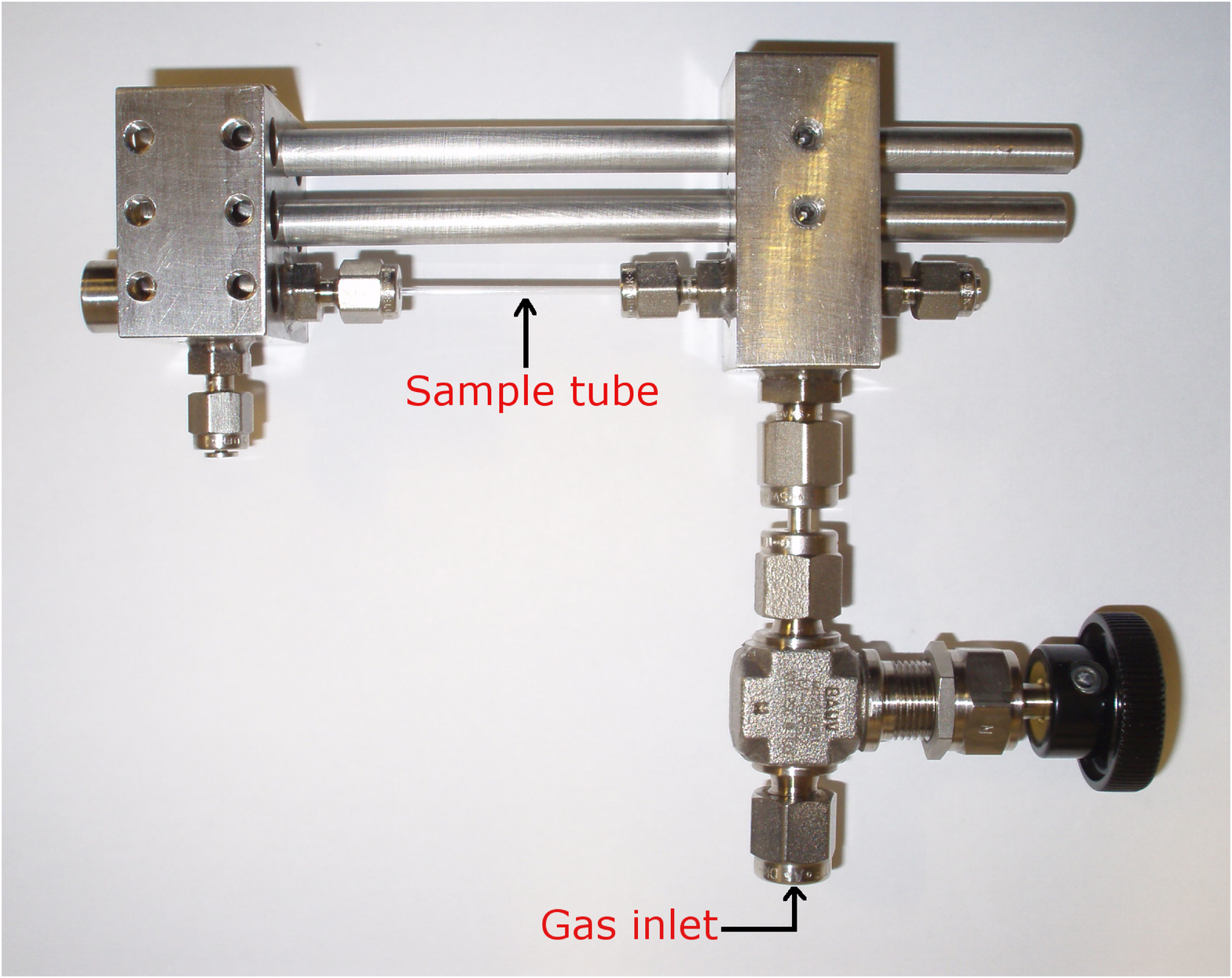}}
\put(-0.3,-2.0){\includegraphics{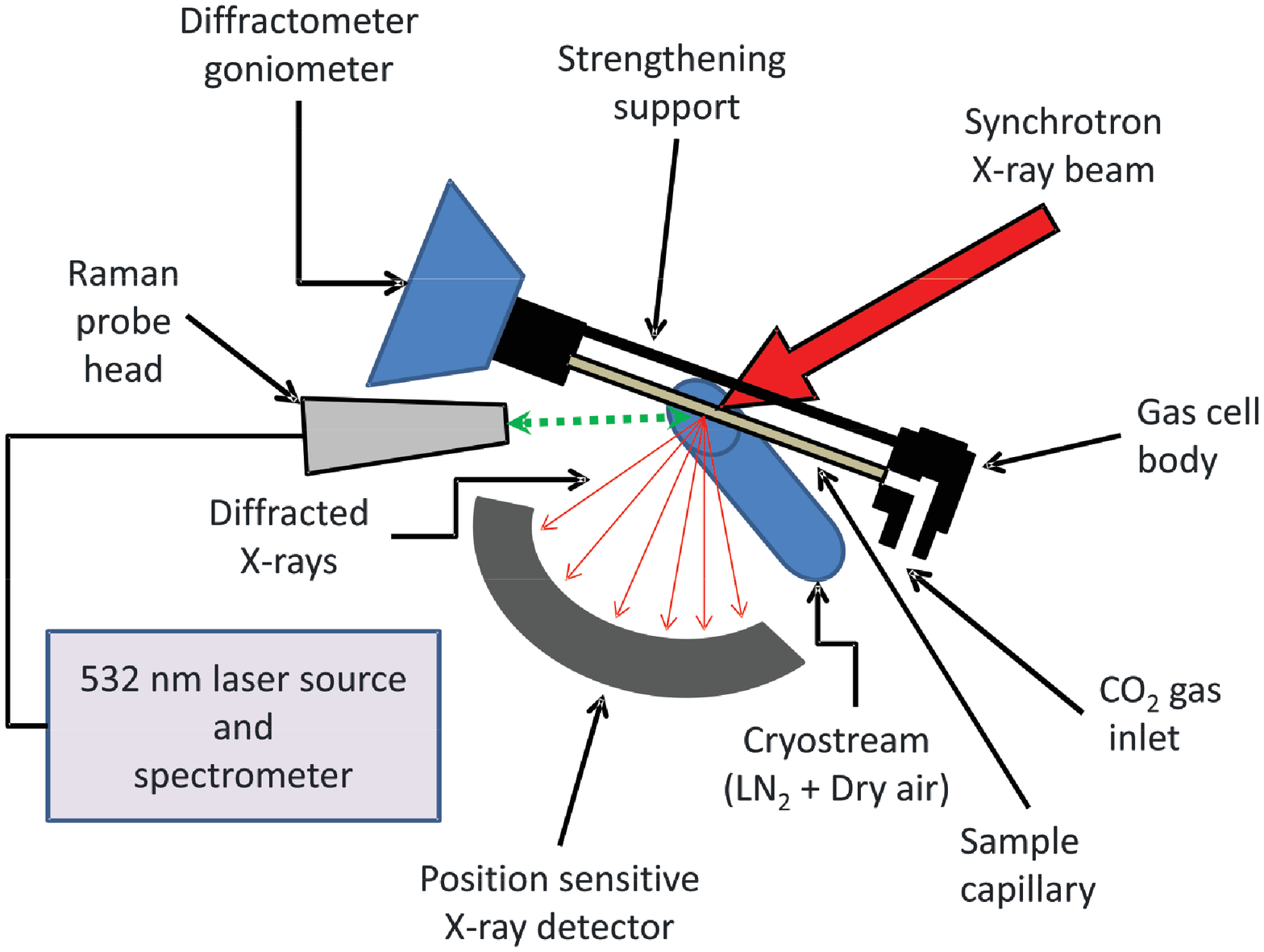}}
\end{picture}
\caption{{\bf (a):} High-pressure cell used to produce clathrates by injecting gas into ice formed by the in situ freezing of water.  {\bf (b):} Experimental arrangement for collecting in situ, variable temperature
SXPD and Raman spectroscopic data on clathrate hydrates. 
\label{setup}}
\end{figure*}

It is suggested that clathrates play an important role in the distribution and total inventory of Martian volatiles, including methane and noble gases (Xe, Ar, Kr), and that the dissociation of extensive sub-surface clathrate deposits could be responsible for some of the geomorphological features observed on the surface of the planet \citep{chassefiere13,mousis13}. They have similarly been suggested as a mode of depletion for the noble gases in the atmosphere of Titan \citep{mousis09,mousis11}, and as the source of the nitrogen depletion observed in comets \citep{iro}.
Clathrate hydrates are also
believed to be present in large quantities within a sub-surface ocean on
Enceladus, and have been suggested as a possible source for the energetic plumes
emanating from its surface \citep{porco}. Many volatile species have been
detected within these plumes, including CO$_2$, CH$_4$, and N$_2$ --  as have
organics \citep{waite} -- and, with the exception of CO$_2$, these gases have
very poor solubility in liquid water. It is therefore considered likely that they exist
below the surface in the form of clathrate hydrates. The likely role of clathrates in the presence of volatile species and noble gases in the evolution of Mars and its atmosphere has been reviewed by \cite{mousis13}; while a summary of CO$_2$ clathrate hydrates in the context of planetary, satellite
and cometary environments is given by \cite{dartois}, who presented 
1.96 -- 4.43 $\mu$m Infrared (IR) spectra at a number of temperatures between 5.6 K and 150 K for CO$_{2}$ clathrates initially formed at 255 K and 20 bar CO$_{2}$ pressure \citep[see also][]{oancea}. \citeauthor{dartois} note the potential diff\-iculties associated with carrying out in situ IR spectroscopy of clathrates in view of their likely location on planetary etc. surfaces and sub-surface regions and their near-inaccessibility to landers. Raman spectra for CO$_2$ clathrates at high pressures (representing clathrate deposits deep within the interior of these icy bodies) have been obtained by \cite{bollengier13} over the 0 -- 1.7~GPa and 250 --330~K pressure and temperature ranges.

Based on thermodynamic calculations, numerous models have been developed for the stability of clathrate hydrates on Europa \citep{prieto05,hand06}, Enceladus \citep{fortes07}, and Titan \citep{mousis11}, as well as comets \citep{marboeuf10}; however, despite clathrate hydrates having a long history of investigation, the available experimental data at the relevant temperatures and pressures necessary to confirm the accuracy, or otherwise, of these models is still limited. \citet{ambuehl14} recently performed detailed kinetic studies of CO$_2$ clathrate hydrate formation/dissociation over a temperature and pressure range relevant to the surface of Mars. However, this experiment was based solely on bulk gas consumption measurements and therefore contributions from microscale structural processes at different temperature and pressure conditions are not taken into account. Understanding the fundamental structural changes that occur during the formation and subsequent break down of the hydrate cages is essential to the understanding of clathrates in different environments and their utility for gas sequestration. \cite{choukroun10} used Raman spectroscopy, coupled with a liquid nitrogen cooled sapphire anvil cell, to study the stability of methane clathrates in relation to the outgassing of methane on the surface of Titan covering 240 -- 320~K and 0 -- 800~bar. However, operationally the setup of the sapphire anvil cell and initial clathrate formation is a difficult multi-step process. 
Other in situ studies of clathrate hydrates have been performed using synchrotron X-ray diffraction \citep{koh96,takeya00a,uchida03} and neutron diffraction \citep{henning00,koh00}. However, these studies have predominantly focused on the stability regimes of clathrate hydrates in relation to terrestrial clathrate deposits e.g. hydrate occurances in gas and oil pipelines and in Arctic permafrost. The temperature and pressure ranges of these experiments are not directly comparable to solar system environments. In addition, kinetic studies show that, while complete conversion of ice to clathrate is a slow process, depending on temperature and pressure, the initial rate of formation can be rapid \citep[a few minutes;][]{ambuehl14,falenty13,gainey12}. Similarly, the dissociation of clathrates can occur over very short timescales \citep{gainey12}. To understand the structural process governing clathrate formation fast data collection times (few s) are thus required. Previous studies presenting time-resolved data have been limited by collection times of 200s \citep{koh96} or more \citep[300s -- 900s;][]{uchida03,takeya00a,henning00}. 

Using high-brightness synchrotron X-ray radiation and a fast detector we are able, using the in situ cell, to acquire time-resolved structural data typically on a scale of $\sim$ 20 seconds or less, allowing the formation and dissociation processes of clathrate hydrates, at variable temperatures, to be observed in situ. In this note we describe the apparatus used for, and present the results of, test measurements showing its suitability for studying clathrate hydrate evolution over temperature and pressure ranges relevant to many solar system bodies. 

\begin{figure}
\setlength{\unitlength}{1cm}
\begin{picture}(12.0,6.3)
\put(-0.3,-2.0){\includegraphics{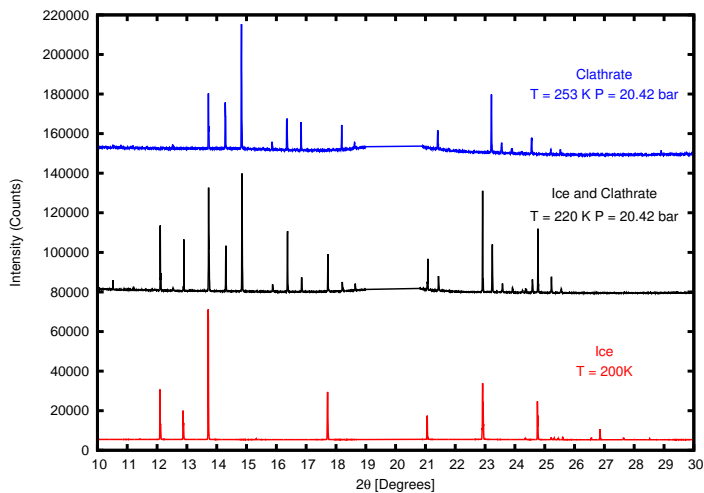}}
\end{picture}
\caption{X-ray powder diffraction patterns for ice at 200~K in air at ambient
pressure (bottom), ice and CO$_2$ clathrate mixture at 220~K and CO$_2$ pressure
20.42~bar (middle) and pure CO$_2$ clathrate at 253~K and CO$_2$ pressure
20.42~bar (top). The strong reflection at $\sim20^\circ~2\theta$ due to the
sapphire capillary has been removed from the data. \label{diff}}  
\end{figure}

\section{Experimental}

The experimental work was carried out at Diamond Light Source Beamline I11 \citep{thompson-I11}. The high pressure sample cell employs $\frac{1}{16}$" Swagelok fittings, mounted into a stainless steel body with bracing supports (Fig.~\ref{setup}a). Deionised water (18.2 M$\Omega$) was loaded into a 0.8~mm diameter single-crystal sapphire tube which is then sealed into the cell by vespel ferrules drilled to match the diameter of the tube. The sample cell is mounted horizontally onto the central $\theta$ circle of a vertical  concentric three circle diffractometer. The sample tube is aligned with the instrument's centre of rotation using a goniometer mount and the gas cell itself is connected to a gas delivery system containing a turbo pump allowing for multiple gas dosing/evacuation cycles \citep{parker}. Sample cooling is provided by a LN$_2$ Cryostream (Oxford Cryosystems), and X-ray diffraction data were collected in the azimuthal plane using the beamline's $90^\circ$ arc position sensitive detector \citep{thompson-psd}. A schematic of the experimental arrangement is shown in Fig. 1b. Using the gas cell and LN$_2$ Cryostream it is in principle possible to simulate a wide range of temperature and pressure conditions (80~K -- 500~K, 1--100~bar), such as those found in solar system environments (e.g. at the surface of Titan and at depth within the sub-surface ocean on Enceladus and Europa).  The in situ Raman system consists of a 532nm laser, Raman probe with a long working length objective lens (65mm working distance), mounted on a motorised linear drive located on a sample table next to the diffractometer and connected to a iHR550 imaging spectrometer. For the present work the X-ray wavelength was 0.826411\AA, calibrated against NIST SRM640c standard Si powder and the beam size at the sample was  2.5~mm (horizontal) $\times$ 0.8~mm (vertical). To improve sampling statistics and compensate for preferred crystal orientation effects resulting from the starting solution being frozen in situ, the sample cell was rocked $\pm 15^{\circ}$ about its length using the diffractometer's $\theta$-circle motion during diffraction data collection. The standard angular speed for this circle on the I11 diffractometer is 4$^{\circ}$ s$^{-1}$, and (compensating for acceleration and deceleration at either end of the rock) requires $\sim$10 s to complete one full $\pm$15$^{\circ}$ rotation. To compensate for the small gaps between the 18 Si-strip detector modules that comprise the I11 position sensitive detector, two 10 s exposures were made with the detector position offset by 0.25$^{\circ}$.  Data collection times were 20~s for the X-ray diffraction and 10-15~s for Raman. Owing to space limitations at the sample point, simultaneous SXPD and Raman measurements cannot be performed, instead the Raman probe is remotely driven in and out of position as required, allowing Raman spectra to be collected between SXPD scans. Therefore for the present work, SXPD was used to monitor ice/clathrate formation for temperature-CO$_2$ pressure combinations and complementary Raman spectra up to 5000~cm$^{-1}$ were obtained at crucial points of interest, though the reverse procedure could equally be followed.

The sapphire tubes, loaded with deionised water, were mounted in the gas cell and aligned with the X-ray
beam. The conversion of ice to clathrate is a temperature-dependent process
\citep{henning00} and once frozen at 200~K, forming a hexagonal ice phase (Ih), the ice was exposed to high purity CO$_2$ (99.98\%) at 20.4 bar. Following the gas injection, conversion of the ice to clathrate
was achieved by increasing the temperature towards the melting point of ice;
at 253~K the ice diffraction features disappeared. The sample was then refrozen by
lowering the temperature to 200~K. The formation of the clathrate is immediately evident from
the growth of multiple features (14 -- 19$^{\circ}$, 21.4$^{\circ}$, 23 -- 24$^{\circ}$, 24.6$^{\circ}$, and 25.2$^{\circ}$ 2$\theta$; see
Fig.~\ref{diff}).  Clathrate features were first observed at a temperature of 220~K. For the purpose of demonstrating the cell's capability, once clathrates had formed, the temperature was cycled between 200~K -- 280~K, to determine the
effect that repetitive cycling would have on the clathrate structure. Finally,
the temperature was held at 270~K while the pressure was steadily decreased in order to observe their dissociation.

\begin{figure}
\setlength{\unitlength}{1cm}
\begin{picture}(10,6)
\put(-0.3,0.0){\includegraphics{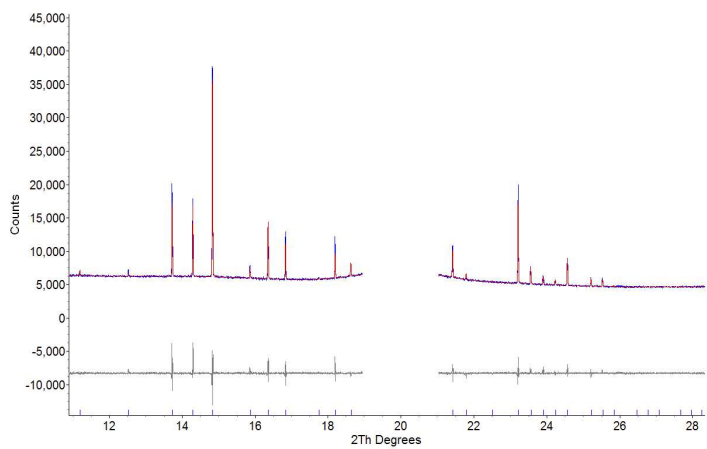}}
\end{picture}
\caption{Rietveld refinement produces a good fit to the experimental data. The experimental data are shown in blue, the calculated fit in red and the residuals in grey below. The strong reflection between $19 - 21^\circ~2\theta$, due to the sapphire capillary, has been excluded from the refinement. \label{refinement}}  
\end{figure}

\section{Results}

\subsection{SXPD}

Figure~\ref{diff} provides a comparison of the X-ray diffraction patterns for pure Ih, mixed ice and CO$_2$ clathrate, and pure CO$_2$ clathrate, for the
conditions specified in the figure caption. Rietveld structure
refinement \citep{rietveld}, using TOPAS refinement software (Version 4.2, Bruker AXS), of the uppermost scan in
Fig.~\ref{diff} (the clathrate at $T=253$~K and $P=20.42$~bar) confirmed the
clathrate phase to be cubic, with a lattice parameter of 11.964 $\pm$ 0.001\AA\ and unit cell volume of 1712.8(4) \AA$^3$; this is consistent
with previously reported sI clathrate structures at this temperature \citep[e.g.][]{udachin,takeya10}. A good fit was obtained, with agreement factors of R$_{wp}$ = 2.82 \%  and R$_{exp}$ = 1.31 \%, producing a goodness of fit value of 2.15.  Figure~\ref{refinement} shows the fit to the experimental data.

\subsection{Raman spectroscopy}

\begin{figure*}
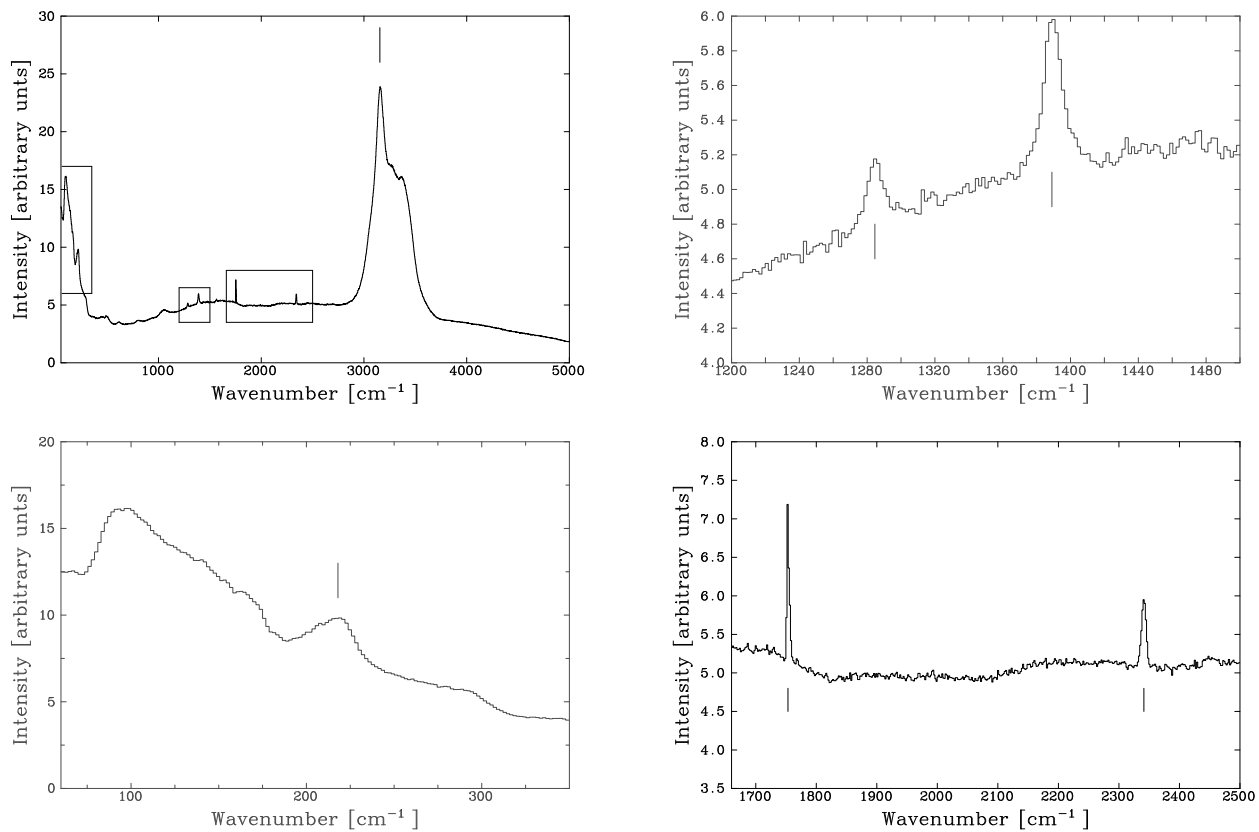

\setlength{\unitlength}{1cm}
\begin{picture}(10.0,11)
\put(-0.3,-2.0){\includegraphics{RAMAN0.eps}}
\put(-0.3,-2.0){\includegraphics{RAMAN1.eps}}
\put(-0.3,-2.0){\includegraphics{RAMAN2.eps}}
\put(-0.3,-2.0){\includegraphics{RAMAN3.eps}}		    
\end{picture}
\caption{Top left: Raman spectrum of CO$_2$ clathrate hydrate at 260~K.
Main feature at $\sim3\,200$~cm$^{-1}$ is due to O$-$H stretch.
Boxed regions are expanded in adjacent plots. 
Top right: close-up of main Raman spectrum in the range
1200~cm$^{-1}$ -- 1500~cm$^{-1}$, showing
$\sim1280$~cm$^{-1}$ and $\sim1390$~cm$^{-1}$ features.
Bottom left: close-up in the range 50~cm$^{-1}$ -- 350~cm$^{-1}$, showing
$\sim220$~cm$^{-1}$ feature. Bottom right: close-up in the range
1600~cm$^{-1}$ -- 5500~cm$^{-1}$, showing
$\sim1753$~cm$^{-1}$ and $\sim2341$~cm$^{-1}$ features.
See text for details. \label{raman}}
\end{figure*}

Raman spectra were obtained in situ and Fig.~\ref{raman} shows the main
features indicative of CO$_2$ clathrates; these data were
obtained at 260~K and 20~bar CO$_2$ pressure. The dominant feature at $\sim$3200~cm$^{-1}$ (see Fig.~\ref{raman}, top left) is
due to the well-known O{\chemone}H stretch and varied little between the clathrate and Ih ice phase, when observed over a temperature and pressure range of 220 -- 270~K and 0 -- 20~bar. 
The shape of the O{\chemone}H stretching bands (and inter-molecular hydrogen
bond vibration bands) however, does depend on the structure of the hydrate. For example,  sII clathrates contain twice as many small 5$^{12}$ cages as large 5$^{12}$6$^4$ ones, whereas the ratio of small cages to large cages in sI hydrates is 1:3. The predominance of large cages --- and the
associated higher ratio of hexagons to pentagons --- induce an orientation of the
water molecules and vibrational behaviour in sI clathrate, in the region of 3200~cm$^{-1}$, that is similar to
water molecules hydrogen bonded in ice. On the other hand, the predominance of small cages in
sII clathrates results in a band profile that is more similar to liquid water \citep{sum,ikeda98}. This means that, when taken in conjunction with other features (see below), the Raman spectrum of the H$_2$O molecule is a good indicator for the clathrate structure \citep{schicks}. 

\begin{figure}
\setlength{\unitlength}{0.95cm}
\begin{picture}(12.0,6)
\put(-0.3,-2.0){\includegraphics{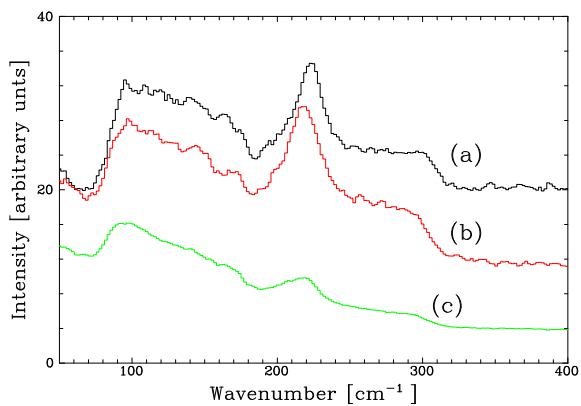}}
\end{picture}
\caption{Raman spectrum of (a) ice at 10~bar CO$_2$ pressure
and temperature 225~K;
(b) ice plus clathrate  at 10~bar CO$_2$ pressure and temperatue 265~K;
(c) CO$_2$ clathrate hydrate at 260~K.
All in the region of the 220~cm$^{-1}$ O$-$O feature. \label{raman0}}
\end{figure}

The feature at $\sim220$~cm$^{-1}$ (see Fig.~\ref{raman}, bottom left) is due to
the inter-molecular O{\chemone}O vibration mode of the water molecules forming the
clathrate structure \citep{nakano}. This feature is shown in greater detail in
Fig.~\ref{raman0} for ice, ice plus clathrate, and clathrate, for which the peak
wavenumbers are $223.1\pm0.5$~cm$^{-1}$, $218.8\pm0.3$~cm$^{-1}$, and
$216.2\pm0.3$~cm$^{-1}$, respectively. These values are consistent
with this feature (unlike the above O{\chemone}H stretch features) being sensitive to environmental changes \citep{nakano}: over the pressure range $\sim100-5000$~bar the Raman shift
varies from $\sim205$~cm$^{-1}$ to $\sim224$~cm$^{-1}$. \citeauthor{nakano} also note that the
220~cm$^{-1}$ O{\chemone}O feature is specific to the CO$_2$ clathrate hydrate;
this can be explained by the structural similarities between the hydrogen
bonded H$_2$O structures of sI clathrate and ice, in that both liquid water and
sII clathrate show only weak broad features in this region.

The features at $\sim1280$~cm$^{-1}$ and $\sim1390$~cm$^{-1}$ (Fig.~\ref{raman}, top right) are the CO$_2$
Fermi diad $\omega_1$:$2\omega_2$ resonance in CO$_2$ molecules trapped
within the clathrate cages; we note that, in pure crystalline CO$_2$ at 6~K,
these features appear at $1275.7\pm0.1$~cm$^{-1}$ and $1384.0\pm0.1$~cm$^{-1}$ ($\Omega_- A^-_g$), and $1276.1\pm0.1$~cm$^{-1}$ and $1384.0\pm0.1$~cm$^{-1}$ ($\Omega_- F^-_g$) \citep{ouillon}. \cite{nakano} found that 
there is little or no pressure-dependence of the Raman shift for the Fermi diad
features, up to a pressure of $\sim5000$~bar. On the other hand, these features
are sensitive to isotopic composition, at least in the
fluid state \citep{irmer,windisch}. In the vapour phase the Raman spectrum of CO$_2$ shows two major, narrow Fermi diad bands
and two minor bands -- denoted as hot bands -- which are coupled through Fermi
resonance.  When CO$_2$ is incorporated into the hydrate lattice, the major
bands are still very pronounced; however, the hot bands convolute into the Fermi
diad bands, contributing only to the asymmetric tails of the bands observed  in
Fig.~\ref{raman}, top right \citep{sum}. No splitting of the diad bands was observed in our data,
showing that the CO$_2$ molecules occupy only the large cavity, as originally
suggested by \cite{ratcliffe}. 

The Fermi diad at $\sim1280$~cm$^{-1}$ and $\sim1390$~cm$^{-1}$, and the
220~cm$^{-1}$ and the 3200~cm$^{-1}$ OH band features can therefore be used as complementary indicators not only
of the presence of clathrate hydrates in astrophysical environments, but (in the case of the Fermi diad bands) also of
C and O isotopic ratios \citep{irmer,windisch} and of the nature of the clathrate environment itself. 


\subsection{Thermal and pressure cycling}

\begin{figure}
\setlength{\unitlength}{1cm}
\begin{picture}(12.0,5.6)
\put(-0.3,-2.0){\includegraphics{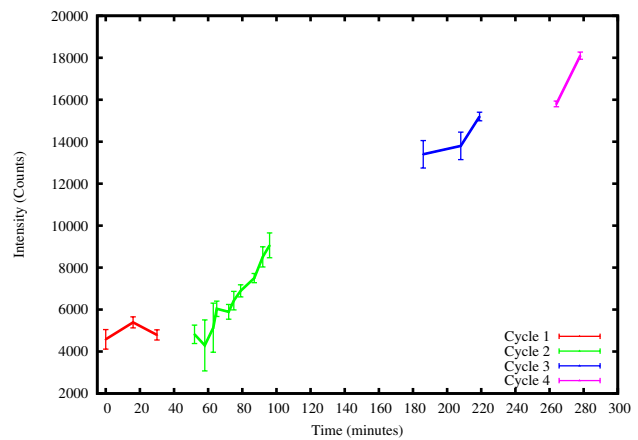}}
\end{picture}
\caption{Effect of thermal cycling on the intensity of the 320 clathrate
peak, seen at 14.3$^{\circ}$ in Fig.~\ref{diff}.
Cycle~1: 200~K -- 245~K. Cycle~2: 210~K -- 230~K. Cycle~3: 220~K -- 240~K.
Cycle~4: 220~K -- 250~K. \label{cycling}} 
\end{figure}

We performed thermal cycling of the clathrates to demonstrate the feasibility of using the gas cell setup to simulate
temperature variations on planetary bodies, such as might occur during day/night
cycles, seasonal variations, or the tidal stressing that occurs on Enceladus
\citep[e.g.][]{hedman}. A total of four cycles were performed, each
within the temperature range 200~K -- 253~K; see Fig.~\ref{cycling} for details.
In this figure the peak intensity of the 320 clathrate X-ray reflection is taken as
a proxy for the formation and destruction of the clathrate, and the abscissa is
given in terms of the time since first cooling to 200~K. A clear and systematic
increase in the peak intensity of the 320 peak with each cycle is observed.
As the sample volume in the beam is unchanged as the temperature is cycled
(apart from negligible thermal expansion effects), this clearly indicates
that, on each cycle, the proportion of CO$_2$ clathrate hydrate is increasing, suggesting that more CO$_2$ is being incorporated into clathrate. A plausible explanation centres on the behaviour of the ice near its melting point. Firstly, as the melting point is 
approached, the ice forms a quasi liquid-like medium surrounding the clathrate \citep{mizuno,henning00}, 
allowing for increased inward diffusion (compared to diffusion through clathrate) of additional CO$_2$ from the environment. Secondly, since measured Raman shifts of the CO$_2$ molecule dissolved in the liquid phase are very close to those of CO$_2$ 
in the clathrate phase the structure of water around CO$_2$ in the liquid phase is likely to be similar to that of the hydrate cage 
\citep{nakano}, suggesting that hydrate formation is easier/faster in the quasi liquid-like phase. Molecular dynamics simulations \citep{jacobson10,jacobson11} of clathrate formation from the liquid phase suggest that crystalline clathrates result from a multi-step process. The first step involves solvent separated clusters of the guest molecules in which the water molecules reorganise to produce polyhedral clathrate cages, resulting in the formation of an amorphous clathrate nucleus. The amorphous nucleus then either reorganises into a crystalline nucleus and grows a crystalline clathrate, or a crystal grows directly around the precursor amorphous seed. These simulations suggest that nucleation of clathrates under conditions of high supercooling, at temperatures well below the melting temperature of the amorphous clathrate, can result in the formation of a metastable amorphous clathrate phase. The ongoing formation/survival of amorphous clathrate precursors, or water molecule cage structures \citep{takeya00b}, could be the origin of the  so-called memory effect reported by certain authors, whereby water that has been previously frozen, or has previously formed hydrate and then thawed, will form hydrate more readily a second time \citep{takeya00b,ohmura03} thus increasing the subsequent rate of formation. 
The forming of a quasi liquid-like medium below the melting point would allow delocalisation of already formed hydrate cages within the ice to easily occur, generating an active surface for forming additional clathrate by allowing delocalised, previously unincorporated and amorphous hydrate, along
with newly formed hydrate cages from the external gas, to attach to the growing clathrate structure. All of these factors contribute to 
maintaining an apparently high ice to (crystalline) clathrate reaction rate \citep{kawamura}. Since the rate of clathrate formation is slower at low temperatures, this would suggest that those otherwise cold solar system objects that experience only brief or intermittent warming events could, nevertheless, over cosmic timescales still build up extensive clathrate deposits that may be a mixture of amorphous and  crystalline phases.

\begin{figure}
\setlength{\unitlength}{1cm}
\begin{picture}(12.0,6.)
\put(-0.3,-2.0){\includegraphics{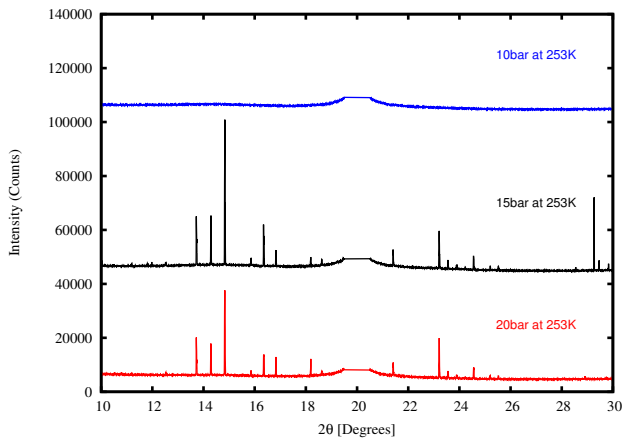}}
\end{picture}
\caption{Pressure-dependence of CO$_2$ clathrate hydrate.
As in Fig.~\ref{diff} a strong feature at $\sim20^\circ~2\theta$ due to the
sapphire capillary has been removed. \label{dissoc}} 
\end{figure}

To demonstrate the effect of a decrease in pressure on the clathrate
structure, we fixed the temperature at 253~K and, starting from a pressure of
20~bar, we reduced the pressure in steps of 5~bar. We found that a reduction in
pressure below 10~bar resulted in rapid ($\ltsimeq20$~mins) dissociation of the
clathrates, as shown in Fig.~\ref{dissoc}. The rapid removal of clathrate as the
pressure falls may be relevant to the production of plumes observed on bodies such as  Enceladus,  where it is proposed that the plumes could result from removal of surface pressure, causing sudden dissociation of clathrates. The sudden dissociation of such sub-surface
clathrate hydrates could be due to a rapid drop in pressure through surface ice fracturing \citep{kieffer}. 

\section{Conclusion}

We have demonstrated the use of a simple, easy to use gas cell for the formation of CO$_2$ clathrates combined with in situ, time-resolved SXPD and in situ Raman spectroscopy. The experimental setup has great potential for studying the formation and stability of clathrate hydrates in astrophysical, particularly solar system, environments. We have presented the results of example measurements to demonstrate its suitability for such work. While the results presented here were performed using pure H$_2$O and only CO$_2$ gas, the cell can be used with various gases (e.g. N$_2$, CH$_4$,
Kr, Xe, and Ar), with multiple gas dosing cycles and different initial liquid compositions (e.g. saline solutions), under a wide range of applied environmental conditions of temperature and pressure (80~K -- 500~K, 1--100~bar), providing a novel experimental resource for the study of clathrate hydrates.

Raman features in the 100 -- 5000~cm$^{-1}$ range are, when taken as a whole, all diagnostic of clathrate formation, while the weak  Raman features at $\sim220$~cm$^{-1}$ are sensitive to changes in environmental conditions. These features could
potentially be used as identifiers of clathrate hydrates or to determine environmental conditions using Raman spectrometers on future planetary landers. 

The formation of clathrates at temperatures $<200$~K was found to be extremely
slow and, within the time frame of the beamtime allocation, could not be
observed. However, we plan to conduct further experiments at these lower temperatures on the recently commissioned Long Duration
Experiment facility on Beamline I11 at Diamond, which allows the in situ processing and monitoring of samples held under slowly varying conditions on a timescale of months to years.

\begin{acknowledgements}
This work was supported by the Diamond Light Source through beamtime award
EE-8037.
SJD acknowledges support from Keele University and Diamond Light Source. The authors would like to extend their thanks to the anonymous referee for their helpful and constructive comments on an earlier version of this paper.
\end{acknowledgements}

\end{document}